\documentclass[12pt]{article}
\usepackage{amssymb,amsmath}
\usepackage{young}
\def\be{\begin{equation}}
\def\ee{\end{equation}}
\def\bea{\begin{eqnarray}}
\def\eea{\end{eqnarray}}
\def\re#1{(\ref{#1})}

\newtheorem{Th}{Theorem}
\newtheorem{Prop}{Proposition}
\newtheorem{Cor}{Corollary}
\newtheorem{Lem}{Lemma}
\newtheorem{Def}{Definition}

\font\largerm=cmr12 at 24pt
\font\cmssl=cmss10 at 12 pt

\newcommand{\bt}{\begin{Th}\ }
\newcommand{\et}{\end{Th}}
\newcommand{\bp}{\begin{Prop}\ }
\newcommand{\ep}{\end{Prop}}
\newcommand{\bc}{\begin{Cor}\ }
\newcommand{\ec}{\end{Cor}}
\newcommand{\bl}{\begin{Lem}\ }
\newcommand{\el}{\end{Lem}}
\newcommand{\bd}{\begin{Def}\ }
\newcommand{\ed}{\end{Def}}
\newcommand{\bconj}{\begin{Conj}\ }
\newcommand{\econj}{\end{Conj}}
\newcommand{\pf}{\noindent{\it Proof:\ }}
\newcommand{\qed}{\hfill $\square$}

\DeclareMathOperator{\sgn}{sgn}
\DeclareMathOperator{\ad}{ad}
\DeclareMathOperator{\Tr}{Tr}

\begin{document}

\begin{titlepage}
\vskip 1 cm
\begin{center}
{\largerm Ternutator Identities}
\end{center}
\vskip 0.6 cm
\begin{center} 
{\cmssl
Chandrashekar Devchand$^1$, David Fairlie$^2$, Jean Nuyts$^3$, Gregor Weingart$^4$}
\vskip 0.6 true cm
{\small  
$^1$ devchand@math.uni-potsdam.de\\ 
{\it   Institut f\"ur Mathematik der Universit\"at Potsdam,\\[1pt]
Am Neuen Palais 10, D-14469 Potsdam, Germany}
\\[3pt]
$^2$ david.fairlie@durham.ac.uk\\
{\it  Department of Mathematical Sciences, University of Durham, \\[1pt]
Science Laboratories, South Rd, Durham DH1 3LE, England}
\\[3pt]
$^3$  jean.nuyts@umh.ac.be\\
{\it    Physique Th\'eorique et Math\'ematique,
Universit\'e de Mons-Hainaut,
\\[1pt]
 20 Place du Parc, B-7000 Mons, Belgium}
\\[3pt]
 $^4$   gw@matcuer.unam.mx\\
{\it   Instituto de Matem\'aticas, Universidad Nacional Aut\'onoma de M\'exico,
\\[1pt]
62210 Cuernavaca, Morelos, Mexico
}}

\end{center}
\vskip 1 cm
\begin{abstract}
The ternary commutator or ternutator, defined as the alternating sum 
of the product of three operators, has recently drawn much attention as 
an interesting structure generalising the commutator. The ternutator 
satisfies cubic identities analogous to the quadratic Jacobi identity
for the commutator. We present various forms of these identities and 
discuss the possibility of using them to define ternary algebras. 
\end{abstract}

\end{titlepage}

\section{Introduction \label{Intro}}

Some time ago Nambu \cite{N} introduced  antisymmetric brackets  depending on three
functions and a derivation,  generalising the Poisson bracket and satisfying the Nambu 
identity, which generalises the Jacobi identity. 
These brackets have been discussed for example in \cite{BF,T,P}.
Nambu also considered the possible quantisation of his bracket:
for three arbitrary associative operators $A_1,A_2,A_3$ in an arbitrary vector space, 
he defined  the related ternary commutator $\left[A_1,A_2,A_3\right]$, which we call {\it ternutator}, by the alternating sum over the permutations of $(1,2,3)$, 
\begin{eqnarray}
 [A_1,A_2,A_3] &:=&  
 A_1A_2A_3 + A_2A_3A_1 + A_3A_1A_2 \nonumber \\
 && - A_1A_3A_2  - A_2A_1A_3 - A_3A_2A_1 \nonumber \\
 &=:& \sum_{\pi\in S_3} \sgn(\pi)  A_{\pi(1)} A_{\pi(2)} A_{\pi(3)} \ .
\label{com3}
\end{eqnarray}
This can be expressed in terms of the commutator as
\bea
 [A_1,A_2,A_3] &=&  A_1 [A_2,A_3] + A_2 [A_3,A_1]  + A_3[A_1,A_2] 
\nonumber \\
             &=&   [A_2,A_3] A_1 +  [A_3,A_1] A_2 + [A_1,A_2] A_3\ . 
\label{com3b}
\eea
The ternutator is trilinear in the operators and has the property of being totally skew-symmetric,
\begin{equation}
[A_1,A_2,A_3] = - [A_2,A_1,A_3] = - [A_3,A_2,A_1] \ ,
\label{skew}
\end{equation}
which leads to the cyclic property,
$
[A_1,A_2,A_3] =  [A_2,A_3,A_1] =  [A_3,A_1,A_2] \ .
$
This bracket, as a particular case of those introduced by Fillipov in \cite{F},
has recently attracted renewed interest (e.g. \cite{CZ,O,BL1,CFZ,CJMFZ,BL2}).

The ternutator generalises the commutator, which is the binary operation
underlying Lie algebras. The Jacobi identity  plays a crucial role in the 
classification of Lie algebras. In this paper we present identities for the ternutator, analogous to the Jacobi identity for the commutator. This problem has also been discussed by others, e.g.  \cite{B,BP}. 
All n-linear brackets defined by the alternating sum of the product of any
odd number $n\ge 5$ of operators have recently been shown \cite{CJM}
to satisfy similar identities. 
Just as the Jacobi identity provides
the defining relations for Lie algebras, we discuss the possibility of using  
the ternutator identities to define ternary algebras.
Some specific examples of sets of structure
constants satisfying the ternutator identity are given in Sect.\ref{ex}.

\section{Ternutators and their identities \label{threeprod}}

The Jacobi identity for commutators, i.e. the equality of the 
two ways of writing the ternutator in terms of the commutator
in \re{com3b}, can be expressed as an alternating sum over
permutations of $(1,2,3)$,
\be
\sum_{\iota\in S_3}  \epsilon_{\iota_1 \iota_2\iota_3}
\left[\left[A_{\iota_1},A_{\iota_2}\right],A_{\iota_3}\right]\  =\ 
2 \sum_{ {\iota\in S_3} \atop {\iota_1 < \iota_2}}
\epsilon_{\iota_1 \iota_2\iota_3}
\left[\left[A_{\iota_1},A_{\iota_2}\right],A_{\iota_3}\right]=0\ ,
\label{Jacobi2}
\ee 
where the second sum is over all permutations $(\iota_1, \iota_2,\iota_3)$ of $(1,2,3)$
satisfying the condition $\iota_1 < \iota_2$ and $\epsilon_{\iota_1 \iota_2\iota_3}$ 
is the completely antisymmetric Levi-Civita symbol.

The following result for the ternutator  \re{com3} is well-known (e.g. \cite{B}):

\bl
There are no identities of second order in the ternutator. 
\el

\pf
The most general second order identity  is
\be
\sum_{ 
       {\iota\in S_5} \atop
        {\iota_1<\iota_2<\iota_3\,,\,\iota_4<\iota_5}
     }
 c(\iota_1,\iota_2,\iota_3)
\left[\left[A_{\iota_1},A_{\iota_2},A_{\iota_3}\right],A_{\iota_4},A_{\iota_5}\right]
=0\ ,
\label{twocom3}
\ee
where the sum is over all permutations $(\iota_1,\dots,\iota_5)$  of $(1,\dots,5)$
satisfying the conditions $\iota_1<\iota_2<\iota_3$ and $\iota_4<\iota_5$.
This has ten summands, multilinear in the five operators $A_i,\ i=1,\dots,5$.
It is easy to check that these ten terms are  
linearly independent. Hence the ten coefficients $c(\iota_1,\iota_2,\iota_3)$ must be zero.

\qed

There are two types of monomials cubic in the ternutator,
both involving seven operators. Label any set of seven operators 
(not necessarily linearly independent)
$A_1,\dots,A_7$  and define 
\bea
T_1(A_1,\dots, A_7)&:=& 
\left[\left[\left[ A_1,A_2,A_3\right],A_4,A_5\right],A_6,A_7\right]
     \label{twotype1a}\\
T_2(A_1,\dots, A_7)&:=& 
\left[\left[A_1,A_2,A_3\right],\left[A_4,A_5,A_6\right],A_7\right]\ ,
\label{twotype2a}
\eea
In virtue of the symmetry properties of the ternutator, 
there are 210 independent monomials of the form
$T_1(A_{\iota_1},A_{\iota_2},A_{\iota_3},A_{\iota_4},A_{\iota_5},A_{\iota_6},A_{\iota_7})$ labeled by 
the permutations of $(\iota_1,\dots,\iota_7)$ of $(1,\dots,7)$ satisfying 
$(\iota_1<\iota_2<\iota_3),\,(\iota_4<\iota_5)$ and $(\iota_6<\iota_7)$, and 70 
monomials of the form 
$T_2(A_{\iota_1},A_{\iota_2},A_{\iota_3},A_{\iota_4},A_{\iota_5},A_{\iota_6},A_{\iota_7})$, with conditions
$(\iota_1<\iota_2<\iota_3),\,(\iota_4<\iota_5<\iota_6),\,(\iota_1<\iota_4)$. 
An identity cubic in ternutators, if it exists, is a linear dependence amongst these 280 terms:
\bea  
&&\hspace*{-1.3cm}
 \sum_{ {\iota\in S_7} \atop 
      {\iota_1<\iota_2<\iota_3\,,\,\iota_4<\iota_5\,,\,\iota_4<\iota_6<\iota_7}}
\hspace*{-1.1cm} c_1(\iota_1,\iota_2,\iota_3,\iota_4,\iota_5)
\ T_1(A_{\iota_1},A_{\iota_2},A_{\iota_3},A_{\iota_4},A_{\iota_5},A_{\iota_6},A_{\iota_7})
     \nonumber\\[12pt]
&&\hspace*{+0cm}
+\hspace{-1 cm}  \sum_{ {\iota\in S_7} \atop
                      {\iota_1<\iota_2<\iota_3\,,\,\iota_1<\iota_4<\iota_5<\iota_6}}
\hspace*{-1 cm} c_2(\iota_1,\iota_2,\iota_3,\iota_7)
\ T_2(A_{\iota_1},A_{\iota_2},A_{\iota_3},A_{\iota_4},A_{\iota_5},A_{\iota_6},A_{\iota_7})\ =\ 0\,,
\label{threecom3a}
\eea
where the sums are over all permutations 
$\iota\in S_7\ ,\ \iota: n \mapsto \iota_n\ ,\  n\in \{1,\dots,7\},$
satisfying the given restrictions.
The following result was also known to Bremner \cite{B}.

\bt
Among any seven  operators there exist precisely seven independent 
identities cubic in the ternutator. 
\et
In order to prove this, we have used REDUCE and independently MAPLE to find 
that the 280 coefficients in \re{threecom3a} are constrained
by 273 independent linear relations leaving a seven parameter space of identities.
Hence there is a basis of seven independent identities, which
generalise the usual Jacobi identity for commutators.

Consider seven arbitrary operators and label them $A_i,\, i=1,\dots,7$. 
Pick out one of them, say $A_7$,
and consider the following alternating sums of $T_1$ and $T_2$,
(skew)symmetric under permutations of the remaining six operators:
\bea
I_1(\iota_1,\dots\iota_6\,;\,7)&=&\hspace{-1cm}
     \sum_{ {{\iota\in S_6} \atop { \iota_1<\iota_2<\iota_3\,,\iota_5<\iota_6 }}}
       \hspace{-0.5cm}  
    \epsilon_{\iota_1 \iota_2 \iota_3 \iota_4 \iota_5 \iota_6}\, 
        [[[A_{\iota_1},A_{\iota_2},A_{\iota_3}],A_{\iota_4},A_{7}],A_{\iota_5},A_{\iota_6}]
     \nonumber\\[8pt]
I_2(\iota_1,\dots\iota_6\,;\,7)&=&\hspace{-1cm} 
     \sum_{ {{\iota\in S_6} \atop { \iota_1<\iota_2<\iota_3\,,\iota_4<\iota_5 }}}
       \hspace{-0.5cm} 
      \epsilon_{\iota_1 \iota_2 \iota_3 \iota_4 \iota_5 \iota_6}\, 
[[A_{\iota_1},A_{\iota_2},A_{\iota_3}],[A_{\iota_4},A_{\iota_5},A_{7}],A_{\iota_6}]\ .\label{invI1I2}
\eea
The identity is then
\be
I=I_1+I_2 \equiv 0\ .
\label{ident}
\ee
An alternative way of writing the identity is as follows. 
Take seven operators $A_a,A_b,A_c,A_d,$ $A_e,A_f,A_g$ and single out $A_g$. 
We can write the identity $I$ as the alternating sum over  all permutations 
of the six indices $(a,b,c,d,e,f)$,
\bea
12\ I&= & \hspace{-0.3cm}
\sum_{S(a,b,c,d,e,f)} \hspace{-0.2cm}  
{\rm sgn}(\pi)\, \biggl(\,
 [[[A_{a},A_{b},A_{c}],A_{d},A_{g}],A_{e},A_{f}]\biggr.
      \nonumber\\
&&\hspace{3 cm} +\, \biggl.[[A_{a},A_{b},A_{c}],[A_{d},A_{e},A_{g}],A_{f}]
    \,\biggr)
\equiv 0 \ ,
\label{idgen}
\eea
where $\sgn(\pi)$ is the sign of the permutation.
The factor $12$ here is the number of times a given term is repeated in 
the sum over all permutations.
The seven independent identities correspond to the seven possibilities of
singling out any one, say $A_g$. 
These seven identities
transform under $S_7$ as the direct sum of the representations associated to
the Young tableaux
\begin{equation}
\begin{array}{c}
\vspace{0cm}\cr
\begin{Young}
  \cr     \cr      \cr     \cr     \cr    \cr     \cr
\end{Young}
\end{array}
\quad\quad\begin{array}{c} 
                   \vspace{-0.5cm}\cr
                    \bigoplus 
          \end{array}\quad\quad
\begin{array}{c} 
          \vspace{-0.5cm}\cr
\begin{Young}  
 &  \cr      \cr     \cr     \cr    \cr     \cr
\end{Young}
\end{array}
\end{equation}
These identities were known to Bremner \cite{B} in a slightly different form. They were indepedently rediscovered in this explicit form by one of us (JN).
There exist further equivalent expressions for these identities. Again singling
out $A_7$, we may write 
\bea
   \lefteqn{3\hskip-35pt\sum_{{\mu\in S_6
    \atop\mu_1<\mu_2<\mu_3,\;\mu_4<\mu_5<\mu_6,\;\mu_1<\mu_4}}
    \hskip-40pt\epsilon_{\mu_1\mu_2\mu_3\mu_4\mu_5\mu_6}\,
     [[A_{\mu_1},A_{\mu_2},A_{\mu_3}],[A_{\mu_4},A_{\mu_5},A_{\mu_6}],A_7]}&&
  \nonumber   \\
   &=&
   \hskip-22pt\sum_{{\mu\in S_7
    \atop\mu_1<\mu_2<\mu_3,\;\mu_4<\mu_5<7,\;\mu_6<\mu_7}}
    \hskip-40pt\epsilon_{\mu_1\mu_2\mu_3\mu_4\mu_5\mu_6\mu_7}\,
     [[[A_{\mu_1},A_{\mu_2},A_{\mu_3}],A_{\mu_4},A_{\mu_5}],A_{\mu_6},A_{\mu_7}]
  \nonumber   \\
   &&
   -\;
   2\hskip-35pt\sum_{{\mu\in S_ 7 
    \atop\mu_1<\mu_2<\mu_3,\;\mu_5=7,\;\mu_6<\mu_7}}
    \hskip-35pt\epsilon_{\mu_1\mu_2\mu_3\mu_4\mu_5\mu_6\mu_7}\,
     [[[A_{\mu_1},A_{\mu_2},A_{\mu_3}],A_{\mu_4},A_{\mu_5}],A_{\mu_6},A_{\mu_7}]
\eea
The left-hand side has 10 summands. In the first term of the right hand side 
there are 150 summands and in the second term 60 summands.
This linear dependence amongst monomials cubic in the ternutator
has altenative  expression:
\bea
   \lefteqn{3\hskip-35pt\sum_{{\mu\in S_6
    \atop\mu_1<\mu_2<\mu_3,\;\mu_4<\mu_5<\mu_6,\;\mu_1<\mu_4}}
    \hskip-40pt\epsilon_{\mu_1\mu_2\mu_3\mu_4\mu_5\mu_6}\,
     [[A_{\mu_1},A_{\mu_2},A_{\mu_3}],[A_{\mu_4},A_{\mu_5},A_{\mu_6}],A_7]}&&
  \nonumber   \\
  &=&
   \hskip-20pt\sum_{{\mu\in S_6 
    \atop\mu_1<\mu_2,\;\mu_3<\mu_4,\;\mu_5<\mu_6}}
    \hskip-30pt\epsilon_{\mu_1\mu_2\mu_3\mu_4\mu_5\mu_6}\,
     [[[A_{\mu_1},A_{\mu_2},A_7],A_{\mu_3},A_{\mu_4}],A_{\mu_5},A_{\mu_6}]
  \nonumber   \\
   &&
 \hskip-5pt   +\;
   \hskip-23pt\sum_{{\mu\in S_6 
    \atop\mu_1<\mu_2<\mu_3,\;\mu_4<\mu_5,}}
    \hskip-25pt\epsilon_{\mu_1\mu_2\mu_3\mu_4\mu_5\mu_6}\,
     [[[A_{\mu_1},A_{\mu_2},A_{\mu_3}],A_{\mu_4},A_{\mu_5}],A_{\mu_6},A_7]
   \nonumber  \\
    &&
 \hskip-5pt     -\;
   2\hskip-25pt\sum_{{\mu\in S_6 
    \atop\mu_1<\mu_2<\mu_3,\;\mu_5<\mu_6}}
    \hskip-25pt\epsilon_{\mu_1\mu_2\mu_3\mu_4\mu_5\mu_6}
     [[[A_{\mu_1},A_{\mu_2},A_{\mu_3}],A_{\mu_4},A_7],A_{\mu_5},A_{\mu_6}]
\eea
Here there are 10 summands in the left-hand side, 
90 terms in the first part of the right-hand side, 
60 terms in the second part and 60 in the third part.

\section{Ternary algebras}

A ternary algebra is a vector space in which
a ternary composition is given by the ``ternutation relations'',
\be
\left[e_i,e_j,e_k\right]=t_{ijk}^{\phantom{ijk}m}e_m\ ,
\label{Lie3}
\ee
where  $e_i$ are basis elements and 
the structure constants $t_{ijk}^{\phantom{ijk}m}$ 
are completely antisymmetric in $i,j,k$ and transform as  $(3,1)$-tensors under
a nonsingular change of basis,
$e'_j=S_j^{\phantom{j}k}e_k\ ,\ \det(S)\neq 0\,$.
Inserting these basis elements in the polynomials $I_1$ and $I_2$ in \re{invI1I2} 
yields cubic polynomials in the structure constants,
\bea
I_1(t\,;\, a,b,c,d,e,f\, ;\, g,q)&=&
\sum_{S(a,b,c,d,e,f)} {\rm sgn}(\pi) \,
 t_{abc}^{\phantom{abc}p}\,
      t_{pdg}^{\phantom{pdg}n}\,
      t_{nef}^{\phantom{nef}q}           \nonumber\\[8pt]
I_2(t\,;\, a,b,c,d,e,f\, ;\, g,q)&=& 
\sum_{S(a,b,c,d,e,f)} {\rm sgn}(\pi) \,
t_{abc}^{\phantom{abc}m}
      t_{deg}^{\phantom{deg}n}
      t_{mnf}^{\phantom{mnf}q}\ ,
\label{i1i2}
\eea
where the sum is over all permutations of $(a,b,c,d,e,f)$.
The ternutator identities \re{idgen} then yield identities for the  
structure constants $t_{ijk}^{\phantom{ijk}m}$,
\be
I(t\,;\, a,b,c,d,e,f\, ;\, g,q) := (I_1 + I_2) (t\,;\, a,b,c,d,e,f\, ;\, g,q) =0
\label{fidentity3}
\ee

Just as a set of structure constants satisfying the Jacobi identities converts a
vector space into a Lie algebra, it would be interesting if the identities \re{fidentity3}
similarly provide necessary and sufficient conditions for a set of structure constants 
$\{t_{ijk}^{\phantom{ijk}m}\}$ to define a ternary 
algebra with relations \re{Lie3}.

The associativity of the multiplication implies the quadratic Jacobi identity for the commutator \re{Jacobi2} as well as the cubic identities  \re{idgen} for the
ternutator. If we lift associativity, both these identities do not hold. For the 
alternative multiplication for the basis elements of the imaginary octonions,
$e_i e_j = - \delta_{ij} + \psi_{ijk} e_k\,$, the left-hand side of the Jacobi identity
\re{Jacobi2} is proportional to the associator 
$(e_i, e_j, e_k )= (e_i e_j) e_k - e_i (e_j e_k) = \varphi_{ijkl} e_l\,$,
where both $\psi_{ijk}$ and $ \varphi_{ijkl}$ are completely antisymmetric.
The ternutator $[e_i, e_j, e_k]$ is also proportional to the  associator
and we have verified by direct calculation using REDUCE that the G$_2$-invariant structure constants of the associator algebra 
$\varphi_{ijkl} = \frac16 \epsilon_{ijklmnp} \psi_{mnp}$ 
with nonzero components 
\begin{equation}
 \varphi_{1234} =\varphi_{1256}  =\varphi_{1357}
=\varphi_{1476}=\varphi_{2376} =\varphi_{2475} =\varphi_{3456} =1\ ,
\label{g2}
\end{equation}
corresponding to the octonion structure constants
$\{\psi_{127} =\psi_{163} =\psi_{154} = \psi_{253}=\psi_{246} 
=\psi_{347} =\psi_{567} = 1\}$
indeed do not satisfy the identities   \re{fidentity3}. 
Instead, the two polynomials in \re{i1i2}, evaluated for $\varphi$ are equal.

Indeed, in dimensions six and seven,  for antisymmetric $t$'s, the values of the
two polynomials in \re{i1i2} are equal, so in these dimensions the identities
 \re{fidentity3} are equivalent to either  $I_1 =0$ or $I_2=0$.
 We have checked explicitly that the equality $I_1 = I_2 \neq 0$  
 also holds for the components of the Spin(7)-invariant 4-form
 \cite{CDFN} defined  by  \re{g2} together with the further nonzero components
$\varphi_{ijk8} = \psi_{ijk}$, namely,
$\varphi_{1278} =\varphi_{1386} =\varphi_{1485} = \varphi_{2385}=\varphi_{2468} 
=\varphi_{3478} =\varphi_{5678} = 1$, and hence this 4-form also does not satisfy the ternutator identity.

Let $\pi$ be a permutation of the four indices $(a,b,c,d)$. The following quadratic
equations in the structure constants
\be
\sum_{\pi\in S(a,b,c,d)} {\rm sgn}(\pi)  
t_{abc}^{\phantom{abc}m} t_{mdg}^{\phantom{mdg}q} = 0
\label{wouldbeid}
\ee
are related to the above-mentioned Nambu identity and 
have been advocated \cite{BL2} as defining conditions for algebras \re{com3}. 
These equations are by no means necessary conditions for the existence of the 
ternary algebra since the ternutator does not satisfy a quadratic identity 
(see Theorem 1).
However, it is readily seen that if \re{wouldbeid} holds then our cubic necessary conditions  
\re{idgen} are satisfied. Hence the relations \re{wouldbeid} 
could be sufficient (but by no means necessary) to guarantee the existence 
of the corresponding ternary algebra.

The Cartan-Killing metric plays a crucial role
in the classification of Lie algebras. There are several tensors, 
constructed from the structure constants $t_{ijk}^{\phantom{ijk}m}$,
which possibly could  play a similar role for ternary algebras.
For Lie algebras, in terms of the adjoint map of element $e_j$ 
with respect to an element $e_i$ defined by $\ad_{e_i} e_j ;= [e_i,e_j]$,
the Cartan-Killing form is given by $g(e_i,e_j) := \Tr (\ad_{e_i} \circ \ad_{e_j} )$.
In terms of the structure constants of the Lie algebra we have
$g_{ij}= f_{im}^{\phantom{im}n}  f_{jn}^{\phantom{jn}m}$.
Using the ternutation relations \re{Lie3} to define an analogous endomorphism
of the ternary algebra, 
$X_{e_i,e_j} e_k := \left[e_i,e_j,e_k\right]$, yields the most natural candidate
for a `metric', namely $ \Tr (X_{e_i,e_j} \circ X_{e_k,e_l} )$, which is clearly
antisymmetric in $ij$, antisymmetric in $kl$ and symmetric under interchange of these
ordered pairs (in other words, it has the symmetries of the Riemann tensor). 
This object has components
\be
g_{ij,kl}=t_{ijn}^{\phantom{ijk}m}\,t_{klm}^{\phantom{ijk}n}
\label{metric2form}
\ee
and can serve as a metric on the $N(N{-}1)/2$ dimensional space of 
2-forms indexed by $\{ij\}, \ i<j$.
If this metric is invertible, its inverse
$g^{ij,kl}$ then satisfies
\be
\sum_{k<l} g_{ij,kl}\,g^{kl,mn}=\delta_i^m\delta_j^n
\label{inverse2metric}
\ee
and affords the construction of the further symmetric tensor,
\be
g^{pq}=\frac{1}{2}\left( t_{ijk}^{{\phantom{ijk}}p}g^{ij,kq}
                        + t_{ijk}^{{\phantom{ijk}}q}g^{ij,kp}   \right),
\label{metric}
\ee
which is a candidate for a metric in the linear space of the operators.

If a Lie algebra is semi-simple, a particularly important
subset of Lie algebras, it is well-known that the metric 
$g_{ij}=f_{ia}^{{\phantom{ia}}b}f_{jb}^{{\phantom{jb}}a}$ formed 
from the structure constants is nonsingular and that 
$f_{ijk}=g_{ka}f_{ij}^{{\phantom{ji}}a}$ is completely antisymmetric in its
indices. It would be remarkable if analogous statements hold for ternary algebras. 
In particular we expect a special role for 
ternary algebras having the following two properties: 
\begin{description}
\item{a)}
there exists a basis in the space of operators such
that the structure constants are completely antisymmetric,
\be
\left[e_i,e_j,e_k\right]=t_{ijkm}e_m\ , \quad {\rm{with\ }}t_{ijkm}{\rm{\ antisymmetric}},
\label{antitern}
\ee
\item{b)}
in this basis, a metric appears in the form
\be
\sum_{a<b<c}t_{iabc}t_{jabc}=\lambda \delta_{ij} \ .
\label{antiterndelta}
\ee
\end{description}

\section {Examples}
\label{ex}

The conditions \re{fidentity3} are so overdetermined that a complete 
classification of solutions, without imposing further conditions, 
seems to be a rather difficult task. However, isolated solutions
in lower dimensions can certainly be found. We display four solutions
in six dimensions with the structure constants being given by components
of a four-form $t_{ijkm}$, which satisfy the identities  \re{fidentity3}. 
It is convenient to use the Hodge-dual  two form $\widetilde t\, {=} *t$
with components $
{\widetilde{t}}_{a_1a_2}=\frac1{4!}\epsilon_{a_1a_2a_3a_4a_5a_6}t_{a_3a_4a_5a_6}\,.
$
\vskip 12pt

\noindent
{\bf Example\ 1}
\bea
{\widetilde{t}}_{12} =
-{\widetilde{t}}_{13} =
-{\widetilde{t}}_{14} =
{\widetilde{t}}_{15} =
-{\widetilde{t}}_{23} =
-{\widetilde{t}}_{24} =&&
   \nonumber\\
={\widetilde{t}}_{26} =
{\widetilde{t}}_{35} =
-{\widetilde{t}}_{36} =
{\widetilde{t}}_{45} =
-{\widetilde{t}}_{46} =
{\widetilde{t}}_{56} &=& 1
\label{examp1}
\eea
{\bf Example\ 2}
\be
-{\widetilde{t}}_{12} =
{\widetilde{t}}_{13} =
{\widetilde{t}}_{14} =
-{\widetilde{t}}_{15} 
={\widetilde{t}}_{26} =
-{\widetilde{t}}_{36} =
-{\widetilde{t}}_{46} =
{\widetilde{t}}_{56} = 1
\label{examp2}
\ee
{\bf Example\ 3}
\bea
{\widetilde{t}}_{14} =
{\widetilde{t}}_{15} =
-{\widetilde{t}}_{16} =
-{\widetilde{t}}_{24} 
-{\widetilde{t}}_{25} =
{\widetilde{t}}_{26} =
{\widetilde{t}}_{34} =
{\widetilde{t}}_{35} =
-{\widetilde{t}}_{36} = 1
\label{examp3}
\eea
{\bf Example\ 4}
\bea 
    &&{\widetilde{t}}_{46}= -1,\quad
    {\widetilde{t}}_{56}= 1,
    \nonumber\\
    &&{\widetilde{t}}_{36}= x_1,\quad
    {\widetilde{t}}_{26}= -x_3,\quad
    {\widetilde{t}}_{35}= -x_2,\quad
    {\widetilde{t}}_{34}= x_2,
    \nonumber\\
    &&{\widetilde{t}}_{25}= x_4,\quad
    {\widetilde{t}}_{24}= -x_4,\quad
    {\widetilde{t}}_{23}= x_4 x_1 - x_2 x_3
\label{examp4}
\eea
In the first three examples, the structure constants satisfy the calibration
criterion
\be
t_{abcd}^3=t_{abcd}\quad \Longrightarrow\quad
 {\widetilde{t}}_{ab}^3={\widetilde{t}}_{ab}\ .
\label{calibration}
\ee
Example 4 does not seem to be a calibration.

\section{Conclusions \label{conclusions}}

Ternutators and ternary algebras generalise the familiar
commutators and Lie algebras and have recently appeared in various
contexts in the literature.  In this
article, we have obtained  explicit forms of the
seven identities, cubic in ternutators, which are obeyed
by any seven associative operators. We have shown that these identities
exhaust the space of identities for seven operators. It remains an open
question whether higher order ($q\geq 4$) identities involving $2q+1$  
operators are trivial consequences of the seven basic identities.
The seven independent identities for seven operators clearly imply identities
(of  fourth order in the ternutator) for 
nine operators. As is clear from \re{idgen} the former seven identities can be 
considered as operations $X(B_1,\dots,B_6\,;\,B_7)$, depending on seven operators, skewsymmetric in the first six. 
Amongst the identities for any given choice of nine operators $A_i\,,\, i=1\dots 9$ there
exist those having one of the following structures:
$
A=  [X(A_1,\dots,A_6\,;\,A_7),A_8,A_9]\ ,\quad
B= X(A_1,\dots,A_6\,;\,[A_7,A_8,A_9]),
\\
C=  X(A_1,\dots,A_5,[A_6,A_8,A_9]\,;\,A_7) . $
There are $7{\cdot} {9\choose 2}= 252$ identities of type A, and ${9\choose 3}= 84$
identities of type B and type C. So altogether 420 identities amongst 9 operators
have their origins in the identities for seven operators. 
Are there any further identities? 

As a consequences of these identities, the structure
constants of ternary algebras must obey cubic relations.
which are highly over-determined. We have given some examples
of solutions of these identities in six dimensions.

\section*{Acknowledgements}
Two of us (JN,GW) acknowledge partial funding from the SFB 647 ``Raum-Zeit-Materie''
of the Deutsche Forschungsgemeinschaft for a visit to Potsdam. JN also thanks the Belgian FNRS for support and  Winfried Neun of the Konrad-Zuse-Zentrum f\"ur Informationstechnik Berlin (ZIB) for help with REDUCE.

\end{document}